\documentclass[american,english,preprints,article,accept,moreauthors,pdftex]{mdpi}
\usepackage[T1]{fontenc}
\usepackage[utf8]{inputenc}
\usepackage{array}
\usepackage{booktabs}
\usepackage{textcomp}
\usepackage{multirow}
\usepackage{amsmath}
\usepackage{graphicx}

\makeatletter

\Title{Frictionless motion of diffuse interfaces by sharp phase-field modeling}

\TitleCitation{Frictionless motion of diffuse interfaces by SPFM}

\Author{Michael Fleck$^{1,*}$\orcidA{}, Felix Schleifer$^{1}$\orcidB{}
and Patrick Zimbrod$^{2}$\orcidC{}}

\AuthorCitation{Fleck, M.; Schleifer, F.; Zimbrod, P.}

\address{$^{1}$\quad{}Metals and Alloys, University of Bayreuth, Prof.-Rüdiger-Bormann-Straße
1, 95447 Bayreuth, Bavaria, Germany\\
$^{2}$\quad{}Applied Computer Science, University of Augsburg, Am
Technologiezentrum 8, 86163 Augsburg, Bavaria, Germany}

\corres{Correspondence: michael.fleck@uni-bayreuth.de}

\abstract{Diffuse interface descriptions offer many advantages for the modeling
of microstructure evolution. However, the numerical representation
of moving diffuse interfaces on discrete numerical grids involves
spurious grid friction, which limits the overall performance of the
model in many respects. Interestingly, this intricate and detrimental
effect can be overcome in Finite Difference (FD) and Fast Fourier
Transformation (FFT) based implementations by employing the so-called
Sharp Phase-Field Method (SPFM). The key idea is to restore the discretization
induced broken Translational Invariance (TI) in the discrete phase-field
equation by using analytic properties of the equilibrium interface
profile. We proof that this method can indeed eliminate spurious grid
friction in the three dimensional space.   Focussing on homogeneous
driving forces, we quantitatively evaluate the impact of spurious
grid friction on the overall operational performance of different
phase-field models. We show that the SPFM provides superior degrees
of interface isotropy with respect to energy and kinetics. The latter
property enables the frictionless motion of arbitrarily oriented diffuse
interfaces on a fixed 3D grid.}


\providecommand{\tabularnewline}{\\}

\firstpage{1} 
\pubvolume{1}
\issuenum{1}
\articlenumber{0}
\pubyear{2022}
\copyrightyear{2022}
\datereceived{} 
\dateaccepted{} 
\datepublished{} 
\hreflink{https://doi.org/10.48550/ \linebreak arXiv.1910.05180} % If needed use \linebreak
\doinum{10.48550/arXiv.1910.05180}
\usepackage{tabularx}
\usepackage{textgreek}

\newcommand*\patchAmsMathEnvironmentForLineno[1]{%
  \expandafter\let\csname old#1\expandafter\endcsname\csname #1\endcsname
  \expandafter\let\csname oldend#1\expandafter\endcsname\csname end#1\endcsname
  \renewenvironment{#1}%
     {\linenomath\csname old#1\endcsname}%
     {\csname oldend#1\endcsname\endlinenomath}}% 
\newcommand*\patchBothAmsMathEnvironmentsForLineno[1]{%
  \patchAmsMathEnvironmentForLineno{#1}%
  \patchAmsMathEnvironmentForLineno{#1*}}%
\AtBeginDocument{%
\patchBothAmsMathEnvironmentsForLineno{equation}%
\patchBothAmsMathEnvironmentsForLineno{align}%
\patchBothAmsMathEnvironmentsForLineno{flalign}%
\patchBothAmsMathEnvironmentsForLineno{alignat}%
\patchBothAmsMathEnvironmentsForLineno{gather}%
\patchBothAmsMathEnvironmentsForLineno{multline}%
}

\newcommand{\mv}[1]{\mathbf{#1}}

\newcommand{\width}{\lambda}
\newcommand{\intEnergy}{\Gamma}
\newcommand{\kin}{M}
\newcommand{\dx}{\Delta x}
\newcommand{\dt}{\Delta t}
\newcommand{\df}{\mu}
\newcommand{\interp}{h}
\newcommand{\gridCoup}{a} % grid coupling parameter
\newcommand{\EngFunc}{F}
\newcommand{\engDens}{f}

\newcommand{\ponderation}{\gamma}

\newcommand{\grid}{\mathbf{p}}
\newcommand{\nbs}{j}
\newcommand{\dir}{k}
\newcommand{\bound}{b}

\newcommand{\gvec}{\mathbf{r}_{\dir}}
\definecolor{link_color}{HTML}{00406E} % dark blue
\definecolor{cite_color}{HTML}{590000} % dark red
\definecolor{dark-violet}{HTML}{9400d3} 
\definecolor{forest-green}{HTML}{228b22} 
\definecolor{dark-red}{HTML}{8b0000} 
\definecolor{dark-blue}{HTML}{00008b} 
\definecolor{dark-pink}{HTML}{ff1493}
\definecolor{dark-yellow}{HTML}{c8c800}
\definecolor{dark-orange}{HTML}{c04000}
\definecolor{dark-salmon}{HTML}{e9967a}
\definecolor{midnight-blue}{HTML}{191970}
\definecolor{dark-spring-green}{HTML}{008040}
\definecolor{newcolor}{rgb}{.8,.349,.1}

\makeatother

\usepackage{babel}
\begin{document}

\section{Introduction}

\label{sec:Introduction}Diffuse interface descriptions, such as phase-field
models, provide an elegant way of modeling microstructure evolution
involving phase or domain boundary motion. In these models the diffuse
interfaces serve as \textquotedbl smeared out\textquotedbl{} volumetric
surrogates for surface-type defects. The surface-type defects are
carriers of physical access energy and their motion of is driven by
the Gibbs-Thompson effect of reducing the total amount surface energy
as well as other volumetric driving forces. As compared to sharp interface
descriptions, the difficult problem of explicit surface tracking is
avoided, which allows for any topological evolution of the phase or
domain structures, such as interface instabilities, shape bifurcations,
nonlinear pattern selection, particle nucleation or dissolution. Phase-field
methods are extensively used in the simulation of complex microstructure
evolution problems, such as solidification \citep{KurzRappazTrievedi2021PartII,TourretLiuLLorca2021,WangBoussinotBrenerSpatschek2021},
solid-state transformations \citep{AstaBeckermKarma012009,WangLi012010,JokisaariNaghaviWolvertonVoorheesHeinonen2017,MianroodiShanthrajKontisCormierGaultSvendsenRaabe2019,MianroodiShanthrajSvendsenRaabe2021},
crack propagation \citep{PonsKarma2010,ChenBouchbiKarma082017,LubomirskyChenKarmaBouchbinder2018,MesgarnejadKarma2020},
ferro-electric domain evolution \citep{YadavNguyenChenSalahuddin2019},
grain growth \citep{MoelansBlanpaiWollant072008,DarvishiKamachaliAbbondandoloSiburgSteinbach2015,DimokratiLeBouarBenyoucefFinel2020},
as well as many other \citep{NiYuJiangHe2017,TonksAagesen2019,AagesenSchwenTonksZhang2019,KimShermanAagesenVoorhees2020,GranasyRatkaiTothGilbertPupaPusztai2021}.

Quite often, the width of the diffuse interface appears to be the
smallest physical length-scale in the system. Clearly, in order to
increase the numerical efficiency in all these cases, one is interested
in choosing the smallest possible numerical width resolution whilst
still keeping the benefits from the diffuse interface description.
So far, spurious grid friction or grid pinning in the phase-field
equation has been the major limiting factor in this regard.  
\begin{figure}
\begin{raggedright}
\begin{adjustwidth}{-\extralength}{0cm}\hspace{1.5cm}%
\begin{minipage}[t][1\totalheight][c]{0.6\textwidth}%
\input{./Figures/Introduction/Pinning-plot.dtex}%
\end{minipage}\hspace{0.5cm} %
\begin{minipage}[t][1\totalheight][c]{0.48\textwidth}%
\input{./Figures/Introduction/Pinning-plot-zwei.dtex}

\input{./Figures/Introduction/Pinning-plot-drei.dtex}%
\end{minipage} \end{adjustwidth} 
\par\end{raggedright}
\caption{\label{fig:pinning}Illustration of the influence from spurious grid
friction on the motion of a constantly driven, planar interface. a)
Comparison of conventional Continuum Field (CF) models for different
phase-field profile resolutions $\tilde{\width}=\width/\dx$ with
the Translationally Invariant (TI) model for $\tilde{\width}=0.5$
(green curves) b) The energy density and the interface velocity, as
measured during the simulation, is plotted as function of the advancing
interface center $\tilde{c}_{n}(t)=c_{n}(t)/\dx$. The dimensionless
driving force is $\tilde{\df}=\df\dx/\intEnergy=0.1$. The CF-model
is subject to pinning for the case of $\tilde{\width}=2.0$. The logarithmic
scale bar on the right shows how much smaller the relative velocity
error for the TI-model is in comparison to those from the CF-model.
 An animated version of this figure is provided in the supplementary
material.}
\end{figure}

For an understanding of what spurious grid friction is, consider an
interface between two phases at different bulk free energy density
levels. The level difference, also called the driving force, induces
an interface motion lowering the total free energy of the system by
lowering the volume of the high energy phase. After some transient
period of time, a homogeneous and time independent driving force should
always result in a stationary state with a constant transformation
velocity. The resulting stationary interface velocity $\upsilon$
is proportional to the driving force, and is exactly prescribed by
energy conservation principles. 

Fig.~\ref{fig:pinning} illustrates the influence from spurious grid
friction on the stationary interface motion.  An animated version
of this figure is provided in the supplementary material. We compare
the resulting interface motion for different dimensionless profile
resolutions $\tilde{\width}=\width/\dx$, i.e.~the ratio between
the phase-field profile width and the numerical grid spacing. In Fig.~\ref{fig:pinning}a)
the phase-field values at different grid points (full symbols) as
well as a least square fit of the profile function Eq.~(\ref{eq:phase-field-tanh-profile-solution})
around the interface region are plotted. On the right, in Fig.~\ref{fig:pinning}b)
the total interface energy and the fitting value for the phase-field
profile width are plotted both as function of the dimensionless position
of the interface center $\tilde{c}_{n}(t)=c_{n}(t)/\dx$, which takes
an integer value whenever a grid point is located in the middle of
the interface. For conventional Continuum Field (CF) type phase-field
models the interface propagates with a clearly smaller average velocity
than expected. This indicates a spurious, friction-like loss of energy
during the interface motion. Further, we obtain oscillations in the
interface energy and velocity as the interface center passes one grid
point after the other. With decreasing profile resolution, we obtain
increasingly larger drops of the average values as well as increasing
oscillation amplitudes. For the CF-model this culminates in fully
destroyed interface kinetics at the profile resolution $\tilde{\width}=2.0$
and below, which is commonly referred to as grid pinning, see yellow
curves in Fig.~\ref{fig:pinning}.  Grid friction and pinning during
stationary interface motion in phase-field modeling has been studied
earlier \citep{KarmaRappel041998,BoeschHMKShochet1995}. Please note,
that the coupling of the phase-field to a local bulk energy density
difference is prototypical for many advanced phase-field models.
In modeling of microstructure evolution such inhomogeneous driving
forces are calculated from local temperatures \citep{WangSekerkaWheelerMurrayCoriellBraunMcFadden1993,Kobayashi1993,AbelBrenerHMK1997},
local concentrations \citep{EchebarFolchKarma122004,EikenBottgerSteinba062006,AagesenGaoSchwenAhmed2018,WangBoussinotHueterBrenerSpatschek2020,ZimbrodSchilp2021},
local strains \citep{FleckPilipenSpatsch122010,BhadakSankarasubramanianChoudhury2018,MesgarnejadPanErbShefelbineKarma2020}
or combinations of these \citep{Steinba2009,SteinbachReview2013,CotturaLeBouarAppolairFinel2015,SongTourretMotaPeredaBilliaTrivediKarma2018}. 

Recently, Finel et al.~found a strikingly novel and surprisingly
simple way to deal with spurious grid friction in one dimension \citep{FinelLeBouarDabasAppolairYamada2018}.
 The method is conceptually related to previous suggestions to improve
the numerical performance of phase-field solvers based on the phase-field
profile function \citep{Glasner2001,Weiser2009,DebierreGuerinKassner2016,GongChenCaoKanLi2018,ShenXuYang2019,JIMolaviTabriziKarma2022,SakaneTakakiAoki2022}.
Similar formulations involving the section wise defined sinus like
phase-field profile have been independently proposed by J.~Eiken
\citep{Eiken2012}.  The key is to restore the Translational Invariance
(TI) in the discretized phase-field equation by using analytical properties
of the phase-field profile function, see Fig~\ref{fig:pinning} green
curve. The TI-model, as shown by the green-curves in Fig.~\ref{fig:pinning},
is neither subject to grid friction nor to grid pinning even-though
the phase-field width has been chosen as small as $\width/\dx=0.5$.
Note, that choosing $\width/\dx=0.5$ means that over $96.4\:\%$
of the hyperbolic tangent interface profile is resolved by just one
grid point, see Eq.~(\ref{eq:phase-field-tanh-profile-solution}).

\begin{figure}
\begin{centering}
\includegraphics[width=1\columnwidth]{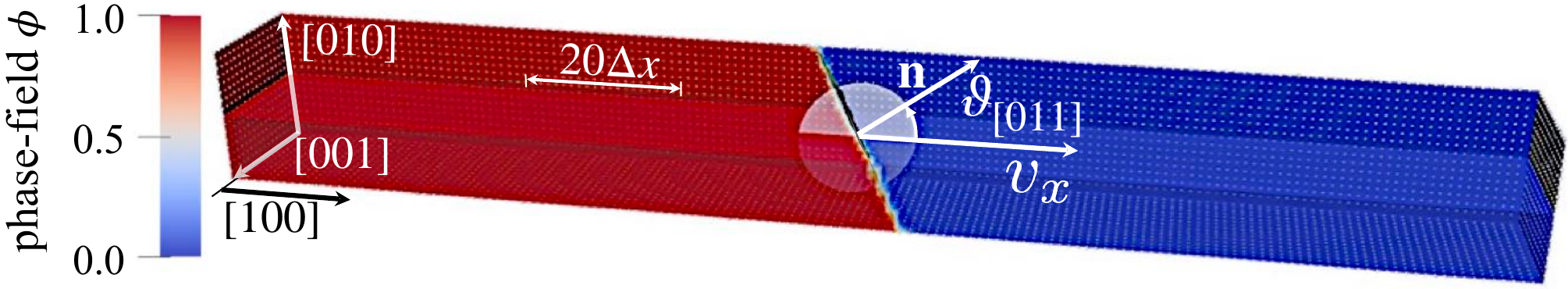}
\par\end{centering}
\caption{\label{fig:Simulation-configuration}The simulation of stationary
interface motion with interface orientations, that differ from the
principal axes of the computational grid.}
\end{figure}

The aim of this work is to proof  that spurious grid friction can
be eliminated by the sharp phase-field method in one dimension as
well as in three dimensions. We define test configurations, which
allow the comprehensive, quantitative evaluation of the intricate
influence from grid friction effects on the operational performance
of different phase-field models within a unified Finite Difference
(DF) framework. The starting point is the stationary interface motion
in one dimension driven by a constant chemical potential density difference.
Depending on modeling details, spurious grid friction and pinning
can seriously limit the total parameter range of reasonable spatial
resolution. In the three dimensional case, the interface orientation
relative to the orientation of the computational grid appears as an
additional degree of freedom. Keeping the uniform, cubic computational
grid as fixed, we investigate the influence of varying interface orientations
on the stationary motion, see Fig.~\ref{fig:Simulation-configuration}.
The realization of this configuration requires to impose special boundary
conditions for the phase-field, which meets the boundary plains under
angles that are different from $0^{{^\circ}}$ or $90^{{^\circ}}$.
Within this article, we newly propose suitable boundary conditions
for this purpose. Deviations of the resulting interface velocity from
the theoretic expectations reveal the effect of spurious grid friction.
We show that the SPFM can also provide frictionless motion of planar
interfaces for arbitrary interface orientation, even in the case of
the very low numerical resolution of the diffuse interface profile.
Finally, we discuss possible grid friction effects on the 3D shape
evolution of a single particle within a matrix phase at constant particle
phase volume.

The article is structured as follows: The theoretic aspects of grid
friction and how the sharp phase-field method deals with it is presented
in Sec.~\ref{sec:Sharp-Phase-Field-Modeling}. This is followed by
the description of the newly proposed contact angle boundary conditions
for the phase-field in Sec.~\ref{subsec:Contact-angle-boundary-condition}.
In Sec.~\ref{subsec:Measure-of-the-interface-position}, we discuss
a new method for the accurate local measure of the interface center
and profile width by a nonlinear profile interpolation. The presentation
and discussion of the results can be found in Sec.~\ref{sec:Results-and-discussion}.
It is started, with the one dimensional study of the effect of grid
friction and grid pinning on the operational limits of different phase-field
models, in Sec.~\ref{subsec:Grid-friction}. In Sec.~\ref{sec:Frictionless-motion-in-3D},
we discuss the case of stationary interface motion with nontrivial
interface orientations in the three dimensional space with the aim
to quantitatively evaluate the residual kinetic grid anisotropy of
different phase-field models. Finally, Sec.~\ref{subsec:Interface-energy-anisotropy}
is devoted to the quantitatively evaluation of the residual energetic
grid anisotropy of different phase-field models, by considering the
particle shape evolution toward quasi equilibrium condition at constant
phase volumes. 

\section{Methods}

\label{sec:Methods}

\subsection{The Sharp Phase-Field Method (SPFM)}

\label{sec:Sharp-Phase-Field-Modeling} We begin by going through
the necessary discretization methods in order to implement the SPFM
computationally. The discretized simulation domain in three dimensions
consists of a uniform, cartesian grid of cubic shape, as is usual
for simulations based on the Finite Difference Method where one is
restricted to operating on equispaced, orthogonal grids. We describe
orientations and directions relative to the simple cubic computational
grid using a Miller index notation system, where the three primitive
translation vectors, i.e.~$\left\langle 100\right\rangle $, conveniently
correspond to the systems orthonormal cartesian basis vectors. The
Sharp Phase-Field Method is based on a discrete Helmholtz free energy
functional $F\left[\phi_{\grid}\right]=\sum_{\grid}\engDens_{\grid}\dx^{3}$,
with a grid spacing $\dx$ along the principle axes. The discrete
Helmholtz free energy density $\engDens_{\grid}$ associated to the
grid point $\grid$ is given by
\begin{align}
\engDens_{\grid}= & \frac{\intEnergy}{C_{\intEnergy}\width}\sum_{\dir}\ponderation_{\nbs}\nu_{\nbs}\Big(\frac{\width^{2}}{2}(\partial_{\dir}^{+}\phi_{\grid})^{2}+g_{\dir}(\phi_{\grid})\Big)+\df\,\interp(\phi_{\grid}).\label{eq:SPFM-Interface-energy-density-3D}
\end{align}
where the discrete directional phase-field derivatives, $\partial_{\dir}^{+}\phi_{\grid}$,
are approximated by forward differencing $\partial_{\dir}^{+}\phi_{\grid}\equiv(\phi_{\grid+\gvec}-\phi_{\grid})/\left|\gvec\right|$
and $\gvec$ denotes a numerical grid vector connecting two neighboring
grid points along the direction number $\dir$. Beside the central
grid point $\grid$, the formulation involves grid points on the first
three neighboring shells $\nbs=1,2,3$, with $\left|\gvec\right|_{\nbs}=\sqrt{\nbs}\dx$,
as summarized in Tab.~\ref{tab:The-TI_120-Formulation}. 
\begin{table}
\caption{The three different neighboring shells and all related grid directions
within the simple cubic numerical grid. Exemplary determination of
the grid coupling parameter set for the $\mathrm{TI}_{\left\langle 120\right\rangle }-$formulation\label{tab:The-TI_120-Formulation}.}
\begin{tabular*}{1\textwidth}{@{\extracolsep{\fill}}cccclc}
\toprule 
shell $\nbs$ & $\dir$ & $\grid$ & $-\grid$ & $\left[120\right]\cdot\grid$ & $\left|\mathrm{arctanh}(\gridCoup_{k})\right|$\tabularnewline
\midrule 
\multirow{3}{*}{1} & 0 & $\left[100\right]$ & $\left[\bar{1}00\right]$ & $\left[120\right]\cdot\left[100\right]=1$ & $4/(\sqrt{5}\width)$\tabularnewline
 & 1 & $\left[010\right]$ & $\left[0\bar{1}0\right]$ & $\left[120\right]\cdot\left[010\right]=2$ & $2/(\sqrt{5}\width)$\tabularnewline
 & 2 & $\left[001\right]$ & $\left[00\bar{1}\right]$ & $\left[120\right]\cdot\left[001\right]=0$ & $0$\tabularnewline
\midrule 
\multirow{6}{*}{2} & 3 & $\left[110\right]$ & $\left[\bar{1}\bar{1}0\right]$ & $\left[120\right]\cdot\left[110\right]=3$ & $6/(\sqrt{5}\width)$\tabularnewline
 & 4 & $\left[011\right]$ & $\left[0\bar{1}\bar{1}\right]$ & $\left[120\right]\cdot\left[011\right]=2$ & $4/(\sqrt{5}\width)$\tabularnewline
 & 5 & $\left[101\right]$ & $\left[\bar{1}0\bar{1}\right]$ & $\left[120\right]\cdot\left[101\right]=1$ & $4/(\sqrt{5}\width)$\tabularnewline
 & 6 & $\left[1\bar{1}0\right]$ & $\left[\bar{1}10\right]$ & $\left[120\right]\cdot\left[1\bar{1}0\right]=-1$ & $2/(\sqrt{5}\width)$\tabularnewline
 & 7 & $\left[01\bar{1}\right]$ & $\left[0\bar{1}1\right]$ & $\left[120\right]\cdot\left[01\bar{1}\right]=2$ & $2/(\sqrt{5}\width)$\tabularnewline
 & 8 & $\left[10\bar{1}\right]$ & $\left[\bar{1}01\right]$ & $\left[120\right]\cdot\left[\bar{1}01\right]=-1$ & $2/(\sqrt{5}\width)$\tabularnewline
\midrule 
\multirow{4}{*}{3} & 9 & $\left[111\right]$ & $\left[\bar{1}\bar{1}\bar{1}\right]$ & $\left[120\right]\cdot\left[111\right]=3$ & $6/(\sqrt{5}\width)$\tabularnewline
 & 10 & $\left[\bar{1}11\right]$ & $\left[1\bar{1}\bar{1}\right]$ & $\left[120\right]\cdot\left[\bar{1}11\right]=1$ & $6/(\sqrt{5}\width)$\tabularnewline
 & 11 & $\left[1\bar{1}1\right]$ & $\left[\bar{1}1\bar{1}\right]$ & $\left[120\right]\cdot\left[1\bar{1}1\right]=-1$ & $2/(\sqrt{5}\width)$\tabularnewline
 & 12 & $\left[11\bar{1}\right]$ & $\left[\bar{1}\bar{1}1\right]$ & $\left[120\right]\cdot\left[11\bar{1}\right]=3$ & $2/(\sqrt{5}\width)$\tabularnewline
\bottomrule
\end{tabular*}
\end{table}
 For a given neighboring shell with $m_{\nbs}$ neighboring nodes,
the coefficients $\nu_{\nbs}=3/m_{\nbs}$ correct for the multiplicity
of the shell. Each of the three different summations, i.e.~$\nbs=1:\dir=0\ldots2$,
$\nbs=2:\dir=3\ldots8$ and $\nbs=3:\dir=9\ldots12$, over all the
directions constituting a certain neighboring shell results in an
independent approximation of the continuous phase-field square gradient
contribution to the free energy density. The relative weightings $\ponderation_{\nbs}$
of the three different realizations are chosen to get best possible
energetic isotropy \citep{FinelLeBouarDabasAppolairYamada2018,FleckSchleifer2022a}.
All the equilibrium potentials $g_{\dir}(\phi)$ are minimal at $\phi=0$
and $\phi=1$. These states correspond to the two distinct phases
of the system. $\width$ denotes the width of the diffuse interface,
$\intEnergy$ is the interface energy density, and $C_{\intEnergy}$
is an interface energy calibration parameter. A positive bulk free
energy density difference $\df$ favors the growth of phase $\phi=0$
at the expense of phase $\phi=1$. The interpolation function $\interp(\phi)$
has to satisfy $\interp(0)=0$ and $\interp(1)=1$. Further, a vanishing
slope at $\partial_{\phi}\interp(\phi=0,1)=0$ is demanded, to keep
the local minima of the total potential energy density at $\phi=0$
and $\phi=1$. 

The Allen-Cahn equation prescribes the time evolution of the phase-field
$\partial_{t}\phi_{\grid}$ to be proportional to the functional derivative
of $\EngFunc$ with respect to the phase-field, i.e.~$-\delta_{\phi}\EngFunc$.
We write \citep{FleckFedermannPogorelov2018,FleckMushongPilipen102011}
\begin{align}
3\width\intEnergy\cdot\partial_{t}\phi_{\grid} & =-2\kin\delta_{\phi}\EngFunc,\label{eq:Allen-Cahn-Phase-field-equation}
\end{align}
where $\kin$ is a kinetic coefficient comparable to a diffusion coefficient
with dimension $\left[\kin\right]=\mathrm{m}^{2}\mathrm{s}^{-1}$.
The functional derivative is defined as $\delta_{\phi}F=\partial_{\phi}f_{\grid}-\sum_{\nbs,\dir}\partial_{\dir}^{-}(\partial_{\left(\partial_{\dir}\phi\right)}^{+}f_{\grid}),$
where the second directional derivative, $\partial_{\dir}^{-}$, is
approximated by backward differencing, i.e.~$\partial_{\dir}^{-}\left(\partial f_{\grid}\right)\equiv\left(\partial f_{\grid-\gvec}-\partial f_{\grid}\right)/\left|\gvec\right|$,
and $\partial_{\phi}\equiv\partial/\partial\phi$ abbreviates the
partial phase-field derivative $\phi$. The continuum phase-field
Eq.~(\ref{eq:Allen-Cahn-Phase-field-equation}) promotes solutions
of the form
\begin{align}
\phi_{\mathrm{\grid}} & =\frac{1}{2}\left(1-\tanh\frac{2\left(\grid\cdot\mv{n}-c_{n}\right)}{\width}\right),\label{eq:phase-field-tanh-profile-solution}
\end{align}
where $\mv{n}$ is the unit normal interface vector and $c_{n}=\upsilon_{\mathrm{th}}t$
denotes the central interface position, moving with the velocity $\upsilon_{\mathrm{th}}$.
During the stationary motion of a planar interface, a constant amount
of energy per unit time interval dissipates via the progressing phase
transformation. Thus, total energy conservation dictates the phase
transformation rate as well as the interface velocity to be exactly
determined by the driving force $\df$ via $\upsilon_{\mathrm{th}}=-\kin\df/\intEnergy$.
Note, that the phase-field parameter $\width$, controlling width
of the hyperbolic tangent interface profile, is not uniquely defined
in the literature. Here, $\width$ is defined in such a way that the
fraction of $\tanh2\simeq0.964$ of the total transition from $\phi=0$
to $\phi=1$ happens within the distance of $2\width$ \citep{DimokratiLeBouarBenyoucefFinel2020,FleckSchleifer2022a}.

In equilibrium, i.e.~$\df=0\rightarrow$ $\partial_{t}\phi=0$, the
phase-field Eq.~(\ref{eq:Allen-Cahn-Phase-field-equation}) reduces
to the discrete force equilibrium condition \citep{FleckSchleifer2022a}:
\begin{align}
\sum_{\dir}\ponderation_{\nbs}\nu_{\nbs}\left\{ \frac{\width^{2}}{\gvec^{2}}(\phi_{\grid+\gvec}-2\phi_{\grid}+\phi_{\grid-\gvec})-\partial_{\phi}g_{\dir}(\phi_{\grid})\right\} = & 0.\label{eq:discrete-equilibirum-formulation}
\end{align}
where approximate the laplacian of $\phi$ by combining forward and
backward differencing to the second order central difference formula,
as is usual within the Finite Differrence Method. Note that solutions
obtained from conventional phase-field implementations do not strictly
satisfy the discrete force equilibrium condition. Not even the ideal
solution Eq.~(\ref{eq:phase-field-tanh-profile-solution}) strictly
satisfies Eq.~(\ref{eq:discrete-equilibirum-formulation}), if the
conventional quartic double well potential $g_{\dir}^{\infty}(\phi)=8\phi^{2}\left(1-\phi\right)^{2}/3$
is imposed. Generally, these violations of the discrete force equilibrium
condition become increasingly severe for small profile resolution
numbers. 
\begin{figure}
\begin{raggedright}
\hspace*{0.05\textwidth}\includegraphics[width=0.85\textwidth]{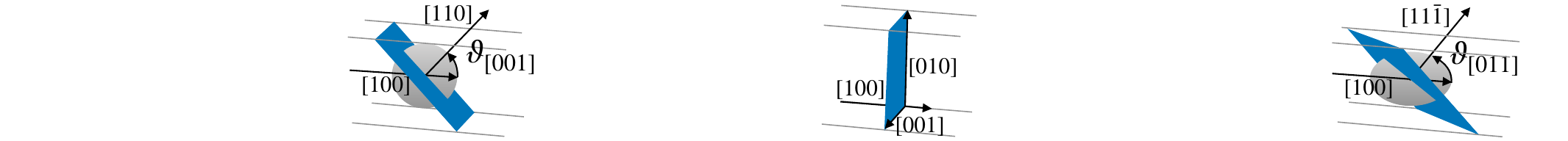}\\
\vspace{-0.2cm}
\begin{flushleft}
\input{./Figures/Methods/pinning_force_vs_orientation.dtex}
\par\end{flushleft}
\par\end{raggedright}
\caption{\label{fig:grid-friction-force-oscillation-vs-Orientation}Logarithmic
plot of the oscillation amplitude $A$ of the total grid friction
forces, i.e.~the system integral over Eq.~(\ref{eq:discrete-equilibirum-formulation}),
using the ideal profile function (\ref{eq:phase-field-tanh-profile-solution})
as a function of the interface orientation angles $\vartheta_{[001]}$
and $\vartheta_{[011]}$. Two different profile resolutions are compared:$\tilde{\width}\!=\!1.0$
(yellow curves) and $\tilde{\width}\!=\!3.0$ (red curves). The dash-dotted
curves indicate the Continuum Field (CF) model, where no TI is restored.
The thick solid curves indicate the TI$_{\left\langle 120\right\rangle }$
model, which uses global grid coupling parameters (\ref{eq:grid-coupling-parameters})
to restore TI for interfaces oriented normal to the $\left\langle 120\right\rangle $-directions.
The TI$_{\mv{n}}$-model (thin curves) restores TI locally in the
direction of the local interface normal $\mv{n}$. The system size
is $300\!\times\!1\!\times\!1$.}
\end{figure}

In Fig~\ref{fig:grid-friction-force-oscillation-vs-Orientation},
we evaluate the degree of satisfaction of the discrete force equilibrium
condition with respect to the ideal profile function (\ref{eq:phase-field-tanh-profile-solution})
for different phase-field models. Therefore, a phase-field is initialized
according to the ideal profile function (\ref{eq:phase-field-tanh-profile-solution})

such that the interface is located in the middle of the computational
domain. The total grid friction force acting on the ideal interface
is given by the system integral over (\ref{eq:discrete-equilibirum-formulation}).
While the continuum force integral clearly vanishes, the discrete
force integral typically oscillates, when the ideal profile is moved
on the computational grid. In Fig.~\ref{fig:grid-friction-force-oscillation-vs-Orientation},
we plot the oscillation amplitude $A$ of the discrete interface force
as a function of the interface orientation for different models. The
conventional Continuum Field (CF) model (dash-dotted curves), as obtained
using the quartic double well potential $g_{\dir}^{\infty}(\phi)=8\phi^{2}\left(1-\phi\right)^{2}/3$,
provides quite large equilibrium force oscillations. For $\tilde{\width}=1$
(yellow curves) the force oscillations clearly reach order unity,
which indicates that the model cannot be operated at such a small
profile resolution. The situation changes for substantially larger
profile resolution numbers, as shown exemplarily by the red curves
in Fig.~\ref{fig:grid-friction-force-oscillation-vs-Orientation}
for the resolution $\tilde{\width}=3$.

Interestingly, for a given interface orientation $\mv{n}$, we can
find a modified equilibrium potential, which strictly satisfies the
discrete equilibrium condition Eq.~(\ref{eq:discrete-equilibirum-formulation}),
thus providing zero grid friction forces for arbitrarily small profile
resolution numbers. This modified potential is derived as follows.
The discrete forces equilibrium condition holds, if all $\dir-$directional
components are simultaneously satisfied. One individual $\dir-$component
can be satisfied at any real time during the propagation of the interface,
based on the following addition property, $\phi_{\grid\pm\gvec}=(1\pm\gridCoup_{\dir})\phi_{\grid}/(1\pm(2\phi_{\grid}-1)\gridCoup_{\dir})$,
of the ideal phase-field profile function. Therefore, we have introduced
the grid coupling parameters as 
\begin{align}
\gridCoup_{\dir}\left(\mv{n}\right) & =\tanh\left(2\frac{\gvec\cdot\mv{n}}{\width}\right).\label{eq:grid-coupling-parameters}
\end{align}
Inserting this addition property into the force equilibrium condition,
we obtain the $\dir-$th component of the derivative of the modified
equilibrium potential \citep{FleckSchleifer2022a}.  Further, integration
leads to the $\dir-$th component of the modified equilibrium potential
\begin{align}
g_{\dir}(\phi)\frac{\gvec^{2}}{\width^{2}}=\phi(1\!-\!\phi)\,+\, & \frac{1\!-\!\gridCoup_{\dir}^{2}}{4\gridCoup_{\dir}^{2}}\ln\Big(\frac{1\!-\!\gridCoup_{\dir}^{2}}{1\!-\!\gridCoup_{\dir}^{2}\left(1\!-\!2\phi\right)^{2}}\Big),\label{eq:equilibrium-potential-TI}
\end{align}
which further satisfies $g_{\dir}\left(0\right)=g_{\dir}\left(1\right)=0$
to allow for easy calculation of the total interface energy (\ref{eq:SPFM-Interface-energy-density-3D})
using an arbitrary phase-field by $\EngFunc_{\mathrm{int}}(\phi_{\grid})=\sum_{\grid}\engDens_{\df=0}$
\citep{SchleiferHolzingerLinGlatzelFleck2019,SchleiferFleckHolzingerLinGlatzel_superalloys2020}.
In the continuum limit $\left|\gvec\right|\rightarrow0$, Eq.~(\ref{eq:equilibrium-potential-TI})
converges to the conventional Continuum Field potential, $g_{\dir}^{\infty}=8\phi^{2}\left(1-\phi\right)^{2}$,
as shown in Fig~\ref{fig:energy-potential}. 
\begin{figure}
\begin{centering}
\begin{center}
\input{./Figures/Methods/potential.dtex}
\par\end{center}
\par\end{centering}
\caption{\label{fig:energy-potential} Joint potential energy density, $\engDens_{\mathrm{pot}}\left(\phi\right)\width/\intEnergy=g_{\dir}\left(\phi\right)/C_{\intEnergy}+\df\interp_{3}\left(\phi\right)\width/\intEnergy$,
as a function of the phase-field $\phi$ with vanishing (approx. parabolic
curves) and non-vanishing (approx. sigmoid shaped curves) driving
force, for different values of the dimensionless interface width $\tilde{\width}=\width/\dx$.
The equilibrium potential $g_{\dir}(\phi)$ is given by Eq.~(\ref{eq:equilibrium-potential-TI}),
and $\interp_{3}(\phi)=\phi^{2}\left(3-2\phi\right)$.}
\end{figure}

This new potential strictly eliminates grid friction forces only for
those ideal profiles having interface orientations that properly relate
to the unit normal vector used to construct the set of grid coupling
parameters defined by Eq.~(\ref{eq:grid-coupling-parameters}). In
the last two columns of Tab.~\ref{tab:The-TI_120-Formulation}, we
construct a set of exemplary grid coupling parameters $\gridCoup_{\dir}\left(\mv{u}\right)$
based on the unit vector $\mv{u}=(1,2,0)^{T}/\sqrt{5}$ pointing in
the $\left[120\right]-$direction of the computational grid. For the
usage of the grid coupling parameters within the sum over the equilibrium
potentials Eq.~(\ref{eq:equilibrium-potential-TI}), the order or
the signs are not important. The final potential value is only determined
by the absolute values of the grid coupling parameters, as given in
the last column in Tab.~\ref{tab:The-TI_120-Formulation}. Note that
all unit vectors $\mv{u}$ pointing in one of the crystallographically
equivalent $\left\langle 120\right\rangle -$ grid directions, as
obtained by all possible permutations and negations of the components,
provide the identical final potential value. In this regard, the $\left\langle 120\right\rangle -$construction
has the advantage that it provides the maximum possible number of
$24$ different equivalent lattice directions, since $h=1\neq k=2\neq l=0$.
The resulting sharp phase-field model, which is constructed from this
set of grid coupling parameters, is denoted as the TI$_{\left\langle 120\right\rangle }-$model.
The thick solid curves in Fig.~\ref{fig:grid-friction-force-oscillation-vs-Orientation}
show that the TI$_{\left\langle 120\right\rangle }-$model provides
vanishing force oscillations for those interface orientations that
match any of the equivalent $\left\langle 120\right\rangle -$numerical
lattice directions. However, the grid friction force evaluation in
Fig.~\ref{fig:grid-friction-force-oscillation-vs-Orientation} also
reveals quite narrow interface orientation windows in which the force
oscillation amplitudes are found to be substantially below the level
of the CF-model. This highlights the sensitivity of the method with
respect to interface orientations. 

In \citep{FleckSchleifer2022a}, we propose a sharp phase-field model,
as denoted by the TI$_{\mv{n}}$-model, which uses locally adaptive
grid coupling parameters. These grid coupling parameters are calculated
based on the locally measured interface orientation. Concerning the
grid friction force evaluation shown in Fig.~\ref{fig:grid-friction-force-oscillation-vs-Orientation},
this model (thin curves) provides very small grid friction force oscillations
regardless of the interface orientation and the profile resolution.
This already indicates that the TI$_{\mv{n}}$-model eliminates spurious
grid friction for arbitrarily oriented planar interfaces in 3D. For
a sufficiently accurate determination of the locally adaptive grid
coupling parameters $\gridCoup_{\dir}\left(\mv{n}\right)$ based on
the local interface orientation $\mv{n}$, the reader is referred
to \citep{FleckSchleifer2022a}.

\begin{table}
\caption{\label{tab:Overview-over-the-models}Overview over all the considered
models constructed within the unified Finite Difference (FD) framework.}
\begin{tabular*}{1\textwidth}{@{\extracolsep{\fill}}llll}
\toprule 
\addlinespace[3pt]
model & equilibrium potential & interpolation function & \multicolumn{1}{l}{Calibration}\tabularnewline\addlinespace[3pt]
\midrule
\addlinespace[3pt]
CF$+\interp_{3}$ & \multirow{3}{*}{$g_{\dir}^{\infty}(\phi)=8\phi^{2}\left(1-\phi\right)^{2}/3$} & $\interp_{3}=\phi^{2}(3-2\phi)$ & \multirow{3}{*}{$C_{\intEnergy}^{\mathrm{CF}}$, $\ponderation_{\nbs}^{\mathrm{TI}_{\left\langle hkl\right\rangle }}$}\tabularnewline
CF$+\interp_{5}$ &  & $\interp_{\mathrm{5}}=\phi^{3}(10-15\phi+6\phi^{2})$ & \tabularnewline
CF$+\interp_{\mathrm{Abel}}$ &  & $\interp_{\mathrm{Abel}}=\phi^{2}/(\phi^{2}+\left(1-\phi\right)^{2})$ & \tabularnewline\addlinespace[3pt]
\midrule
TI$_{\left\langle hkl\right\rangle }+\interp_{3}$ & $g_{\dir}:$ Eq.~(\ref{eq:equilibrium-potential-TI}), $\gridCoup_{\dir}(\mv{u})$,
$\mv{u}\parallel\left\langle hkl\right\rangle $ & \multirow{2}{*}{$\interp_{3}=\phi^{2}(3-2\phi)$} & $C_{\intEnergy}^{\mathrm{TI}}$, $\ponderation_{\nbs}^{\mathrm{TI}_{\left\langle hkl\right\rangle }}$\tabularnewline
TI$_{\mv{n}}+\interp_{3}$ & $g_{\dir}:$ Eq.~(\ref{eq:equilibrium-potential-TI}), $\gridCoup_{\dir}\left(\mv{n}\right)$ &  & $C_{\intEnergy}^{\mathrm{TI}}$, $\ponderation_{\nbs}^{\mathrm{TI}_{\mv{n}}}$ \tabularnewline\addlinespace[3pt]
\bottomrule
\end{tabular*}
\end{table}

Within this article, we compare the behavior of different phase-field
models with respect to grid friction effects. All these models have
been implemented within a unified Finite Difference (FD) framework.
An overview over all the considered models is given in Tab.~\ref{tab:Overview-over-the-models}.
The models basically differ by their choices for the equilibrium potentials
$g_{\dir}(\phi)$ and for the interpolation function $\interp(\phi)$. 

The Continuum Field (CF) models are obtained in the limit $\lim_{\left|\gvec\right|\rightarrow0}$.
In this limit the equilibrium potentials (\ref{eq:equilibrium-potential-TI})
converge to the classical quartic double-well potential $g^{\infty}(\phi)=8\nu\phi^{2}\left(1-\phi\right)^{2}/3$,
and no Translational Invariance is restored. Imposing this potential,
we obtain finite difference implementations for phase-field models,
that correspond to conventional Allen-Cahn type models using a hyperbolic
tangent profile. Best possible comparability to the present sharp
phase-field models is reached by employing the same interface energy
calibrated 27 grid point approximation of the Laplace operator in
the phase-field equation (\ref{eq:equilibrium-potential-TI}). Different
CF-models result from three different choices for for the interpolation
function: (i) the natural interpolation function $\interp_{3}$, (i)
the most frequently used interpolation function $\interp_{5}$, which
provides infinite phase stability and (iii) the broken rational interpolation
function $\interp_{\mathrm{Abel}}(\phi)=\phi^{2}/(\phi^{2}+(1-\phi)^{2})$.

Translational Invariance (TI) is obtained when the new equilibrium
potential, given by Eq.~(\ref{eq:equilibrium-potential-TI}) is imposed
in conjunction with the natural interpolation function $\interp_{3}$.
When the grid coupling parameters $\gridCoup_{\dir}$ in equilibrium
potentials are set to the fixed set $\gridCoup_{\dir}(\mv{u})=\tanh\left(2\gvec\cdot\mv{u}/\width\right)$,
then Translational Invariance is restored for all $\left\langle hkl\right\rangle -$directions
of the computational grid, with $\mv{u}$ parallel to one of the $\left\langle hkl\right\rangle -$directions.
These models are denoted as TI$_{\left\langle hkl\right\rangle }+\interp_{3}$.
A combination the new equilibrium potentials with other interpolation
functions turns out to be not useful, because the nonequilibrium phase-field
profile alternation destroys the carefully restored Translational
Invariance again. In case of the TI$_{\mv{n}}$-model, the grid coupling
parameters $\gridCoup_{\dir}\left(\mv{n}\right)$ are calculated based
on the locally determined interface normal vector $\mv{n}$ \citep{FleckSchleifer2022a}.
Then, TI is restored locally in the local interface normal direction
$\mv{n}$. 
\begin{figure}
\begin{centering}
{\scriptsize{}}\begin{center}
{\footnotesize{}\input{./Figures/Methods/energy_calibration.dtex}}
\par\end{center}{\scriptsize\par}
\par\end{centering}
\caption{\label{fig:energy-Calibration} Different interface energy calibration
parameters $C_{\intEnergy}(\tilde{\width})$ (solid and dashed green
curves) and ponderation coefficients $\ponderation_{2}(\tilde{\width})$
(solid and dash-dotted violet curves), $\ponderation_{3}(\tilde{\width})$
(solid and dash-dotted blue curves) as functions of the profile resolution
$\tilde{\width}$. $\ponderation_{1}=1-\ponderation_{2}-\ponderation_{3}$}
\end{figure}

All models have been separately calibrated using the procedure discussed
in \citep{FleckSchleifer2022a}. The result of these calibration procedures
is a set of profile resolution dependent calibration parameters $C_{\intEnergy}(\tilde{\width})$
and $\ponderation_{\nbs}(\tilde{\width})$ for each model. However,
not all of these calibration parameter functions turn out to be practically
different. For instance, the energy calibration parameter function
turns out to be equal for all different sharp phase-field model. The
different calibration parameter functions are illustrated in Fig.~\ref{fig:energy-Calibration}
and the association to the different models is explained in the last
column of Tab.~\ref{tab:Overview-over-the-models}.  

\subsection{Contact angle boundary conditions}

\label{subsec:Contact-angle-boundary-condition} The simulation of
interface propagation in directions other than the $\left\langle 100\right\rangle -$directions
of the computational grid requires special boundary conditions for
the phase-field. In these simulations the interface has to meet the
boundaries under a definite contact angle. Here, a boundary condition
for the phase-field, enforcing a given interface orientation angle
$\alpha$ with respect to the boundary plane, is newly proposed and
implemented. Physically the condition can be understood as a wetting
angle of droplets on a surface \citep{BenSaidSelzerNestlerBraunGreinerGarcke2014,DiewaldLautenschlaegerStephanLangenbachKuhnSecklerBungartzHasseMueller2020}.
We use the addition property of the ideal phase-field profile Eq.~(\ref{eq:phase-field-tanh-profile-solution})
to impose the profile shift, $s_{n}=\dx\sin\alpha$, on the boundary.
This shift relates to the angle $\alpha$ and enforces the phase-field
to meet the boundary plane with proper orientation. 
\begin{figure*}
\begin{centering}
\includegraphics[width=0.85\textwidth]{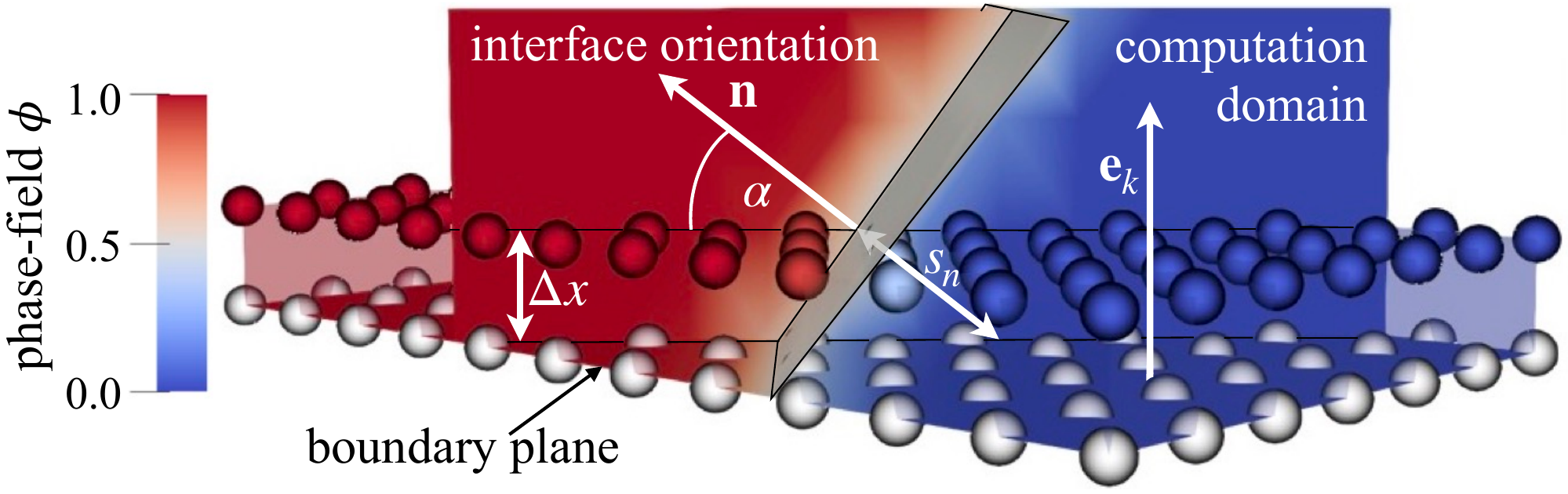}
\par\end{centering}
\caption{Schematic illustration of the boundary conditions for the phase-field
which enforce the wetting or contact angle $\alpha$ between the interface
normal and the direction normal to the boundary plane.\label{fig:Contact-angle-BC}}
\end{figure*}

The case when a phase front with the interface normal $\mv{n}$ meets
a boundary plane with direction $\mv{e}{}_{\dir}$ under an angle
$\alpha$ is shown in Fig.~\ref{fig:Contact-angle-BC}. The interface
is visualized by the gray plane, which relates to the $\phi=1/2-$contour
of the phase-field. The grid points associated with the computational
domain are indicated as colored spheres, with the color denoting the
respective phase-field value. Grid points $\grid_{\bound}$ associated
with the boundary are indicated as gray spheres.  The boundary value
at $\grid_{\bound}$ is calculated as 
\begin{align}
\phi_{\grid_{\bound}} & =\frac{\left(1-\gridCoup_{\dir}\right)\phi_{\grid_{\bound}+\gvec}}{1-\gridCoup_{\dir}\left(2\phi_{\grid_{\bound}+\gvec}-1\right)},\label{eq:wetting-angle-boundary-conditions}
\end{align}
with $\gridCoup_{\dir}=\tanh(2\left|\gvec\right|\sin\alpha/\width)$.
The idea behind Eq.~(\ref{eq:wetting-angle-boundary-conditions})
is to calculate the boundary value at $\grid_{\bound}$ from the neighboring
phase-field value at $\grid_{\bound}+\gvec$ using the addition property
of ideal profile function Eq.~(\ref{eq:phase-field-tanh-profile-solution}),
and imposing a profile-shift by the length $s_{n}=\sin\alpha\left|\gvec\right|$
along the interface normal direction. Experience from the simulations
with different phase-field model has shown that the accuracy of the
proposed contact angle boundary conditions depends on how well the
respective phase-field model reproduces the ideal hyperbolic tangent
profile. 

\subsection{Measure of the interface position and width }

\label{subsec:Measure-of-the-interface-position} A practical way
to accurately determine the actual central interface position $c_{n}$
as well as the current profile width $\width$ for given phase-field
simulation results is to fit ideal profile function (\ref{eq:phase-field-tanh-profile-solution})
to the data using least squares. This could be done, for instance,
using the Marquardt-Levenberg algorithm \citep{FleckSchleifer2022a}.
However, by practical reasons it is not always possible to determine
these quantities in such an elaborate way. In these cases one would
rather like to have an easy and efficient method which, for instance,
just interpolates the central interface position $c_{n}$, i.e.~the
$\phi=0.5-$contour, based on the known positions and phase-field
values at the neighboring grid points. Note that a simple linear interpolation
turns out to be not useful in the present case. The linear interpolation
results are not sufficiently accurate if the phase-field profile width
is small compared to the grid spacing. 

Here, we propose a new nonlinear interpolation technique to calculate
the $l-$contour position based on the analytic phase-field profile
(\ref{eq:phase-field-tanh-profile-solution}). The $l-$contour position
denotes the interpolated position between two neighboring grid points
at the positions $\grid$ and $\grid+\gvec$, separated by the lattice
vector $\gvec$, where the phase-field takes the value $\phi=l$,
with the contour-level $l\in(0,1)$. The two neighboring grid points
are located on opposite sides of the $l-$contour position along the
direction $\dir$, such that the following condition $\left(\phi_{\grid}-l\right)\cdot\left(\phi_{\grid+\gvec}-l\right)\leq0$
is satisfied. Based on the two different phase-field values $\phi_{\grid}$
and $\phi_{\grid+\gvec}$two different $l-$contour positions can
be calculated as 
\begin{align}
x_{\grid}^{\mathrm{int}} & =\mv{e}_{\dir}\cdot\grid+\frac{\hat{\lambda}_{\dir}}{2}\left|\mathsf{arctanh}\frac{l-\phi_{\grid}}{2l\phi_{\grid}-\phi_{\grid}-l}\right|,\label{eq:nonlinear-profile-interpolation-c}\\
x_{\grid+\gvec}^{\mathrm{int}} & =\mv{e}_{\dir}\cdot(\grid+\gvec)-\frac{\hat{\lambda}_{\dir}}{2}\left|\mathsf{arctanh}\frac{l-\phi_{\grid+\gvec}}{2l\phi_{\grid+\gvec}-\phi_{\grid+\gvec}-l}\right|,\label{eq:eq:nonlinear-profile-interpolation-p}
\end{align}
where $\mv{e}_{\dir}$ is a unit vector parallel to the direction
$\dir$ and $\hat{\width}_{\dir}=\width/n_{\dir}$ denotes the directional
phase-field width, as determined by the phase-field parameter $\width$
and the a priorily unknown projection $n_{\dir}$ of the unit normal
interface orientation vector onto the $\dir-$th direction.  Assuming
the two contour positions to be equal $x_{\grid}^{\mathrm{int}}=x_{\grid+\gvec}^{\mathrm{int}}$,
we obtain an estimation for the directional phase-field width 
\begin{align}
\hat{\width}_{\dir} & =2\mv{e}_{\dir}\cdot\gvec\left(\left|\mathsf{arctanh}\frac{l-\phi_{\grid}}{2l\phi_{\grid}-\phi_{\grid}-l}\right|+\left|\mathsf{arctanh}\frac{l-\phi_{\grid+\gvec}}{2l\phi_{\grid+\gvec}-\phi_{\grid+\gvec}-l}\right|\right)^{-1}\label{eq:width-interpolation}
\end{align}
This value for the directional phase-field width is inserted into
Eqs.~(\ref{eq:nonlinear-profile-interpolation-c}) and (\ref{eq:eq:nonlinear-profile-interpolation-p}).
In order to further regularize and symmetrize the finally interpolated
contour position, we impose linear interpolation, $x_{\grid,\gvec}^{\mathrm{int}}=\{x_{\grid+\gvec}^{\mathrm{int}}(l-\phi_{\grid})+x_{\grid}^{\mathrm{int}}(\phi_{\grid+\gvec}-l)\}(\phi_{\grid+\gvec}-\phi_{\grid})$,
between the two slightly different positions at the two neighboring
grid points $x_{\grid}^{\mathrm{int}}$ and $x_{\grid+\gvec}^{\mathrm{int}}$.

\section{Results and discussion}

\label{sec:Results-and-discussion}

\subsection{Frictionless interface motion in 1D}

\label{subsec:Grid-friction} First, we consider the constantly driven
motion of a planar interface in one dimension. After a certain time
that depends on model and profile width, the system reaches a stationary
state of motion where the interface velocity is exactly known from
energy conservation principles and is given as $\upsilon_{\mathrm{th}}=-\kin\df/\intEnergy.$
We perform a simulation study with highly comparable individual simulation
runs, with a constant time resolution of $\kin\df\dt/(\intEnergy\dx)=1.6\cdot10^{-8}$.
In all the individual simulation runs the interface center has passed
at least a minimum number of four grid points (corresponds to $2.5\cdot10^{8}$
time steps) even after the system has reached a stationary state.
To reduce the overall computational demands, the whole system is incrementally
shifted back by one grid point, whenever the fraction of the energetically
favored phase reaches $50\%$ of the system \citep{FleckBrenerSpatsch2010,FleckQuerfurthGlatzel2017}.
Then, it is sufficient to resolve the total system by just $50$ grid
points, which is ten times the maximally employed profile resolution.
Please note, that especially in case of the higher spatial resolution
numbers $\tilde{\width},\tilde{\df}$ some models require very long
transient times to relax from the initial ideal profile to the stationary
state. 

\begin{figure}[t]
\begin{raggedright}
\begin{flushleft}
\input{./Figures/Results/VelocityVsdf-width-1.dtex}\hspace*{-1pt}\input{./Figures/Results/VelocityVsdf-width4.dtex}\hspace*{0.2cm}\input{./Figures/Results/VelocityVsdf-width3.dtex}\hspace*{0.2cm}\input{./Figures/Results/VelocityVsdf-width2.dtex}\hspace*{0.2cm}\input{./Figures/Results/VelocityVsdf-width0.dtex}\hspace*{-1pt}\input{./Figures/Results/VelocityVsdf-width-2.dtex}\\
\vspace{0.2cm}
~\input{./Figures/Results/Evaluation_Parameter_Window_stationary-pf.dtex}
\par\end{flushleft}

\begin{flushleft}
\vspace{-0.2cm}
\centering\hspace*{1cm}dimensionless driving force $\tilde{\df}$
\par\end{flushleft}
\par\end{raggedright}
\caption{\label{fig:Velocity-vs-driving-force}Logarithmic errors of the stationary
interface velocity as a function of the dimensionless driving force
$\tilde{\df}=\df\dx/\intEnergy$ for different profile resolutions:
$\tilde{\width}=\width/\dx=4.0,\;3.0,\;2.5,\;0.5$. The behavior of
different models is compared: (i) Continuum Field (CF) model with
$\interp_{\mathrm{Abel}}$ (violet), (ii) CF model with $\interp_{5}$
(blue), (iii) CF model with $\interp_{3}$ (red) and (iv) the sharp
phase-field model with Translational Invariance (TI$+\interp_{3}$)
(green). Solid lines denote the mean relative errors and the oscillations
are indicated as transparently colored areas (see Fig.~\ref{fig:pinning}).
 The bottom plot compares the parameter windows of reasonable spatial
resolution evaluated for the different models. Here, a model is reasonably
resolved if the mean velocity error is lower than 10\%.}
\end{figure}

In Fig.~\ref{fig:Velocity-vs-driving-force}, we compare average
relative interface velocities as well as the stationary oscillation
amplitudes as function of the constant driving force for a number
of different models. As illustrated in Figs.~\ref{fig:pinning}b),
the oscillation amplitudes are indicated by the transparently colored
areas connected to the solid lines. The colored areas start from the
oscillation amplitude value and end at the mean value. When the colored
area is found above the mean value, we encounter the desirable situation
that the measured value oscillates around the theoretic expectation.
In contrast, colored areas below the mean value denote the undesirable
case when the theoretic expectation is located somewhere outside the
oscillation interval. 

Fig.~\ref{fig:Velocity-vs-driving-force} shows that all Continuum
Field (CF) models are subject to grid pinning for the profile resolution
$\tilde{\width}=2.5$ and dimensionless driving forces below $\tilde{\df}<0.02$.
Due to the absence of any interface motion in these cases, the mean
relative velocity error takes the value $1$ and the measured oscillation
amplitude vanishes, resulting in the large colored areas below the
solid lines on the left side in the logarithmic plots. For increasing
profile resolutions, the onset of pinning is shifted towards smaller
dimensionless driving forces, as also clearly visible in Fig.~\ref{fig:Velocity-vs-driving-force}.
At a profile resolution of $\tilde{\width}=4$, we find a limited
parameter window of driving forces in which the mean relative error
is minimal and nearly constant at a value of about $2\cdot10^{-2}$.
This significant residual error at the comparably large profile resolutions
$\tilde{\width}=4$ indicates the relevance of spurious grid friction
in conventional phase-field modeling. In contrast, the sharp phase-field
model provides very small relative velocity errors and oscillations
amplitudes on the order of the time discretization error for all resolutions
$\tilde{\width},\tilde{\df}<1$. This indicates that spurious grid
friction and grid pinning is truly eliminated in the one dimensional
sharp phase field model. 

We further discuss the application of large dimensionless driving
forces or small interface energy densities. Even-though quite often
disregarded, limitations with respect to large dimensionless driving
forces exist in any phase-field model and denote a relevant restriction
of the general applicability of these models. In all kinds of diffuse
interface descriptions the interface energy area density $\intEnergy$
is somehow distributed over the localized volume covered by the diffuse
interface. Then, the morphological changes due to the interface energy,
i.e.~the so-called capillary forces, are modeled by a volume density
equivalent, which is inversely proportional to the width of the diffuse
interface. The wider the diffuse interface is chosen, the smaller
the volume density equivalent of the interface energy is. Therefore,
large dimensionless driving forces are the natural consequence of
coarse graining or up-scaling of simulations.  An important example
is the simulation of dendritic solidification, which involves small
capillary lengths and large diffusion lengths, resulting in small
dendritic tip radii as well as medium secondary- and large primary
dendrite arm spacings \citep{TourretKarma2016,TourretSturzViardinZalovni2020,FleckQuerfurthGlatzel2017}.
A less complex example is the study of the development of interface
instabilities, such as the diffusional Mullins-Sekerka-\citep{MullinsSekerka1964}
or the elastic Asaro-Tiller-Grinfeld instability \citep{AsaroTiller1972,Grinfeld1993,Kassner2001,SpatschekFleck2007}.
Both require the interface energy to be comparably small.  

Naturally, the choice of the interpolation function is gaining more
importance at large dimensionless driving forces. First, we consider
the case of imposing the natural interpolation function: $\interp_{\mathrm{3}}(\phi)=\phi^{2}(3-2\phi)$
\citep{Kassner2001,PilipenkoFleckEmmerich2011,FinelLeBouarDabasAppolairYamada2018}.
In this case the ideal phase-field profile Eq.~(\ref{eq:phase-field-tanh-profile-solution})
remains an analytic solution of the phase-field Eq.~(\ref{eq:Allen-Cahn-Phase-field-equation})
even at finite driving forces. Then the maximal possible driving
force is given by the condition of phase stability. Phase stability
demands the driving force to be small enough to guarantee meta-stability
of the high energy phase: The two local minima of the potential energy
density at $\phi=0,1$ have to be separated by a (local) maximum in
between. For the CF$+\interp_{3}$-model, phase stability requires
the dimensionless driving force to be below $\tilde{\left|\df\right|}<8/(3C_{\intEnergy}^{\mathrm{CF}}\tilde{\width})$,
with $\tilde{\df}=\df\dx/\intEnergy$ and $\tilde{\width}=\width/\dx$.
The TI$+\interp_{3}$-model provides a profile resolution dependent
phase stability limit, which nicely reflects the behavior of model
\citep{FleckSchleifer2022a}. 

Changing the interpolation function can provide phase stability for
larger driving forces. One example is the interpolation function $\interp_{\mathrm{Abel}}(\phi)=\phi^{2}/(\phi^{2}+(1-\phi)^{2})$,
which has been first proposed by Abel et al.~\citep{AbelBrenerHMK1997}.
The advantage of this interpolation function is that a thermodynamically
consistent extension to the case of multiple phases is comparably
easy \citep{Moelans2011,MoelansBlanpaiWollantPRB072008}. With regard
to the condition of phase stability, we obtain a maximally possible
driving force for this interpolation function of $\left|\tilde{\df}\right|<440/(3C_{\intEnergy}^{\mathrm{CF}}\tilde{\width}).$
An even more common choice for the interpolation function is $\interp_{\mathrm{5}}=\phi^{3}(10-15\phi+6\phi^{2})$
(see e.g.~\citep{Plapp092011,OhnoTakakiShibuta2017,AagesenGaoSchwenAhmed2018,GreenwoodShampurOforiOpokuPinomaaGurevichProvatas2018,KimShermanAagesenVoorhees2020}).
The CF$+\interp_{5}$-model even provides phase stability for infinitely
large driving forces, which is of course a highly desirable property.
However, using interpolation functions other than the natural one
leads to altered nonequilibrium phase-field profiles. 

The profile alternation grows with increasing driving force. Stronger
alternations in turn also lead to stronger grid friction effects.
For large dimensionless driving forces, the two CF-models $\interp_{5}$
and $\interp_{\mathrm{Abel}}$ are both limited by grid friction,
while the two $\interp_{3}$-models are limited by the condition of
phase stability. Neither the use of $\interp_{5}$ nor that of $\interp_{\mathrm{Abel}}$
is useful in the sharp phase-field model, as the altered nonequilibrium
profile destroys any restored Translational Invariance. 

In the lower part of Fig.~\ref{fig:Velocity-vs-driving-force}, the
parameter ranges of reasonable spatial resolution are evaluated for
each of the different models. A model is set to be reasonably spatially
resolved, when the relative velocity error during constantly driven
interface motion is found to be less than $0.1$. While, the CF models
cannot reasonably operate at profile resolutions below $2$, the sharp
phase-field model can. For very small profile resolutions, the sharp
phase-field model provides surprisingly high limits of phase stability
\citep{FleckSchleifer2022a}.

\subsection{Frictionless interface motion in 3D}

\label{sec:Frictionless-motion-in-3D} 
\begin{figure}[t]
\begin{raggedright}
\hspace*{0.08\textwidth}\includegraphics[width=0.85\textwidth]{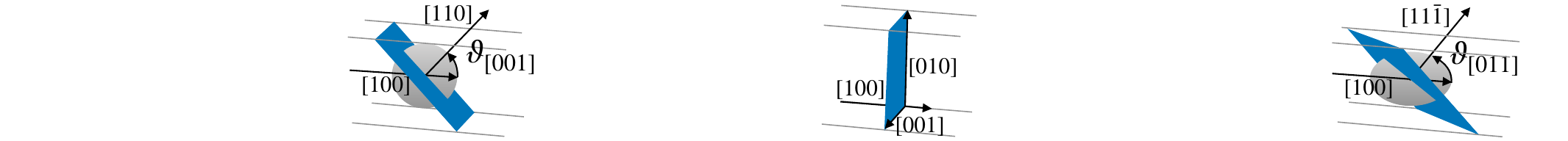}\\
\vspace{-0.3cm}
\begin{flushleft}
\input{./Figures/Results/VelocityVsOrientation-width4.0.dtex}\\
\input{./Figures/Results/VelocityVsOrientation-width1.0.dtex}\\
\input{./Figures/Results/VelocityVsOrientation-width0.5.dtex}
\par\end{flushleft}

\begin{flushleft}
\vspace{-0.2cm}
\centering \hspace*{2cm}interface orientation
\par\end{flushleft}
\par\end{raggedright}
\caption{\label{fig:Velocity-vs-Orientation}Relative error in the $x-$component
of the stationary velocity as functions of the interface orientation
angles $\vartheta_{\left[001\right]}$ and $\vartheta_{\left[011\right]}$.
The results from the Continuum Field (CF)-model, various TI$_{\left\langle hkl\right\rangle }$-models
and the TI$_{\mv{n}}$-model are compared for three different profile
resolutions $\tilde{\width}=4.0$, $\tilde{\width}=1.0$ and $\tilde{\width}=0.5$.
The imposed driving force decreases with increasing orientation angles,
i.e.~$\df=\df_{0}\cos\left(\vartheta\right)$ with $\tilde{\df}{}_{0}=\dx\df{}_{0}/\intEnergy=0.1$
and in any case $\interp_{3}$. The time discretization is $\kin\df_{0}\dt/(\intEnergy\dx)=10^{-6}$.}
\end{figure}

A phase-field model can be
anisotropic with respect to interface kinetics as well as interface
energetics. Here, we investigate both effects separately. The residual
kinetic anisotropy is studied by considering the constantly driven
stationary motion of planar interfaces with varying interface orientations
$\mv{n}$. In a stationary system state, the interface normal velocity
is exactly determined by energy conservation principles $\upsilon_{n}^{\mathrm{th}}=-\kin\df/\intEnergy.$
Within the fixed cartesian grid comprised of $120\times10\times10$
equispaced grid points, differently oriented interfaces meet the cubic
domain boundaries under specific angles. This requires the employment
of the special boundary conditions for the phase-field, as discussed
in section \ref{subsec:Contact-angle-boundary-condition}. The different
interface orientations $\mv{n}$ result from continuous rotations
around the two different axes $\left[001\right]$ and $\left[011\right]$,
as sketched on top of the plot in Fig.~\ref{fig:Velocity-vs-Orientation}.
The respective rotation angles between the interface normal direction
and the $x-$direction, i.e.~the $\left[100\right]-$direction, are
denoted by $\vartheta_{\left[001\right]}$ and $\vartheta_{\left[011\right]}$,
respectively. To evaluate the degree of isotropy of the interface
kinetics, we perform a simulation study consisting of many highly
comparable individual simulations with an equal $x-$component of
the interface velocity $\upsilon_{x}^{\mathrm{th}}$ for all individual
simulations. Therefore, the imposed driving forces are chosen to decrease
with increasing orientation angles, i.e.~$\df=\df_{0}\cos\left(\vartheta\right)$
with $\tilde{\df}{}_{0}=\dx\df{}_{0}/\intEnergy=0.1$, and a time
discretization of $\kin\df_{0}\dt/(\intEnergy\dx)=10^{-6}$. Running
each individual simulation for at least $5\cdot10^{6}$ time steps
ensures that the interface center has passed at least five grid points
along the $\left[100\right]-$direction. 

In Fig.~\ref{fig:Velocity-vs-Orientation}, we compare the orientation
dependent error (mean value as well as oscillation amplitude) in the
interface velocity for different phase-field models. The relative
mean errors are depicted by the solid lines and the relative oscillation
amplitudes are visualized as colored areas, as has been previously
illustrated on the right hand side of Fig.~\ref{fig:pinning}b). 

The black curves shows the results from the Continuum Field (CF) model.
Only for the profile resolution $\tilde{\width}=4$, the CF-model
is not subject to pinning. Note that the vanishing force oscillations
at the profile resolutions $\tilde{\width}=1$ and $\tilde{\width}=0.5$
have been omitted in Fig.~\ref{fig:Velocity-vs-Orientation} for
the CF-model. Even for a profile resolution as large as $\tilde{\width}=4$
the CF-model is still characterized by a significant kinetic anisotropy
of about 3\%, which is on the scale of the mean relative error in
the interface velocity. 

All the sharp phase-field models are more accurate in this regard,
in some configurations even by more than an order of magnitude. For
a profile resolution of $\tilde{\width}=4$, none of these models
is subject to a kinetic anisotropy larger than $0.1\%$. However,
similar to the onset of grid pinning this situation quickly changes
when the profile resolution is decreased. The TI$_{\left\langle hkl\right\rangle }$-models
denote sharp phase-field models with a restored the Translational
Invariance (TI) in the $\left\langle hkl\right\rangle -$directions,
as also discussed in Sec.~\ref{sec:Sharp-Phase-Field-Modeling}.
As expected, for interface orientations close to the directions of
restored translational invariance, we obtain extremely small errors
in the interface velocity. However, already slightly misoriented interfaces
propagate at velocities that are clearly below the expectation. The
resulting errors in the interface velocity indicate the existence
of finite grid friction effects during the stationary interface motion
in these directions. Even spurious grid pinning is observed at the
profile resolutions $\tilde{\width}=1$ and $\tilde{\width}=0.5$
for interface orientations in the vicinity of the $\left[100\right]-$direction
or less pronounced in the $\left[110\right]-$direction for the most
of the TI$_{\left\langle hkl\right\rangle }$-models, as clearly visible
in Fig.~\ref{fig:Velocity-vs-Orientation}.  

The green curves in Fig.~\ref{fig:Velocity-vs-Orientation} show
the behavior of the TI$_{\mv{n}}$-model, where the TI has been locally
restored in the local direction of interface motion. This provides
very accurate interface velocities for all orientations, even if the
phase-field width is chosen to be as small as $\tilde{\width}=0.5$.
The resulting velocity error is basically given by the residual error
in the time discretization.  This observation indicates that this
model indeed provides frictionless motion of planar interfaces with
arbitrary orientations in the three dimensional space.

It should be noted that the TI$_{\left\langle hkl\right\rangle }$-models
differ from the original 3D sharp phase-field model purposed by Finel
in some respects \citep{FinelLeBouarDabasAppolairYamada2018}. We
use a Finite Difference (FD) implementation that operates on a simple
cubic computational grid, whereas the 3D sharp phase-field model is
formulated on a fcc grid using a spectral FFT-based solver. Both aspects
are expected to significantly influence the modeling behavior with
regard to the present investigation on the residual anisotropy profile
of the interface kinetics.

\subsection{Interface energy driven shape relaxation}

\label{subsec:Interface-energy-anisotropy}

\begin{figure}
\begin{raggedright}
\begin{adjustwidth}{-\extralength}{0cm}\begin{center}
\input{./Figures/Results/sphericity_configuration.dtex}{\footnotesize{}\hspace*{0.7cm}}\input{./Figures/Results/eval_sphericity.dtex}
\par\end{center}\end{adjustwidth} 
\par\end{raggedright}
\caption{\label{fig:evaluation-sphericity} Investigation of the residual energetic
anisotropy by interface energy relaxation of an initially cubic particle
at constant particle volume. a) Total interface energy as a function
of the simulation time. b) Anisotropy $\epsilon_{\mathrm{eff}}=\left(R_{\mathrm{max}}-R_{\mathrm{min}}\right)/2R_{\mathrm{mean}}$
of the final quasi equilibrium phase-fields as a function of the phase-field
width $\tilde{\width}$ for different phase-field models. Insets:
Phase-field contours together with their sphericity errors plotted
as color-value. For $\tilde{\lambda}=1.8$ the extended error-range
of is plotted aside the inset.}
\end{figure}
To quantify the residual grid anisotropy of the interfacial energy,
we consider the shape evolution of a single particle towards the quasi
equilibrium state under constant particle volume \citep{BhadakSankarasubramanianChoudhury2018,HolzingerSchleiferGlatzelFleck2019,SchleiferHolzingerLinGlatzelFleck2019,LinSchleiferHolzingerTaSkrotzkiDarvichiKamachaliGlatzelFleck2021}.
The 3D simulations of size $60\times60\times60$ are started with
an initially cubic particle, as shown in the inset of Fig.~\ref{fig:evaluation-sphericity}a).
During the simulation, a homogeneous and time dependent driving force
is imposed, such that the integral volume of the particle neither
shrinks nor grows \citep{NestlerWendlerSelzer072008,FleckMushongPilipen102011,FleckFedermannPogorelov2018}.
The shape evolution of the particle is driven by the minimization
of the total interface energy. Fig.~\ref{fig:evaluation-sphericity}a)
shows the relaxation of the particles total interface energy as a
function of the simulation time during the shape evolution. It rapidly
approaches a distinct shape that reflects the effective interface
energy anisotropy. In case of a fully isotropic model, the resulting
equilibrium shape will be an ideal sphere. Thus, the residual interface
energy anisotropy is defined as the deviation of the resulting equilibrium
shape from the shape of an ideal sphere. 

The residual interface energy anisotropy of a certain phase-field
model operating at a certain profile resolution $\tilde{\width}$
is evaluated from the finally relaxed phase-field at the end of the
respective simulation. The anisotropy is evaluated from nonlinearly
interpolated $\phi=1/2-$contour points. The nonlinear interpolation
of contour positions based on a given phase-field, is described in
section \ref{subsec:Measure-of-the-interface-position}. For each
contour position we calculate its distance from the particles barycenter
$R_{i}$ and divide this by the mean radius $R_{\mathrm{mean}}$.
This ratio is called the local sphericity error and is provided as
the color value in the contour-plots of the quasi equilibrium shapes
shown in Fig.~\ref{fig:evaluation-sphericity}b). Finally, the overall
residual anisotropy of this particular simulation is given by $\epsilon_{\mathrm{eff}}=\left(R_{\mathrm{max}}-R_{\mathrm{min}}\right)/2R_{\mathrm{mean}}$,
where $R_{\mathrm{min}}$ and $R_{\mathrm{max}}$ denote the smallest
and largest distance between the $\phi=1/2-$contour positions and
the particle's barycenter, respectively. 

Fig.~\ref{fig:evaluation-sphericity}b) shows the evaluation of the
residual interface energy anisotropy as a function of the profile
resolution $\tilde{\width}$ for different phase-field models. For
the largest profile resolution $\tilde{\width}=4$ all models show
very small residual anisotropies with sphericity errors below $3\cdot10^{-4}$.
Already at profile resolutions of $\tilde{\width}=2$ and below the
Continuum Field (CF) model provides the highest sphericity errors
of the quasi equilibrium particle shape due to the onset of pinning.
The partially pinned particle contours for the cases $\tilde{\width}=2$
and $\tilde{\width}=1.8$ are exemplarily shown as insets in Fig.~\ref{fig:evaluation-sphericity}b).
For profile resolutions below $1.3$, larger anisotropies due to spurious
grid pinning is also observed using the TI$_{\left\langle hkl\right\rangle }$-models,
especially in the generic cases, when the faces of the initial cube
are not accidentally aligned with the directions of restored TI. The
residual interface energy anisotropies obtained for the TI$_{\mv{n}}$-model
are generally very low but, of course, gradually increase with decreasing
profile resolution, leading to maximal sphericity errors of $2.0\:\%$
for $\tilde{\width}=0.4$ or $3.0\:\%$ for $\tilde{\width}=0.35$.

\section{Conclusion}

\label{subsec:Conclusion} The intricate effects of spurious grid
friction, grid pinning and grid anisotropy limit the performance of
conventional phase-field models in many ways. To large extends, these
limitations can be overcome by the Sharp Phase-Field Method (SPFM)
\citep{FleckSchleifer2022a,FinelLeBouarDabasAppolairYamada2018}.
We quantitatively evaluate the operational limits of different phase-field
models with and without the SPFM within a unified finite difference
framework. An overview of all the different models, considered in
this work, is given in Tab.~\ref{tab:Overview-over-the-models}.
The operational limits of the models are defined as the borders in
the parameter space separating reasonable from erroneous model behavior.
The parameter space of interest is spanned by the dimensionless driving
force $\tilde{\df}=\df\dx/\intEnergy$ and the dimensionless profile
resolution $\tilde{\width}=\width/\dx$. The key results and findings
of this work can be summarized as follows:
\begin{itemize}
\item Spurious grid friction is studied by means of simulations of stationary
interface motion in one dimension, as shown in Fig.~\ref{fig:Velocity-vs-driving-force}.
In the limit of small driving forces all CF-models are limited by
grid pinning, while the sharp phase-field model is entirely free of
this effect. With respect to the important case of large dimensionless
driving forces all models involving the natural interpolation function
$\interp_{3}$ are limited by the condition of phase stability. The
other models are limited by spurious grid friction due to increasingly
stronger alternations of the phase-field profile. 
\item The residual kinetic anisotropy of the models is evaluated by systematic
variations of the interface orientation within the 3D simulation of
constantly driven interface motion. When imposing a one-grid-point
interface resolutions as small as $\tilde{\width}=0.5$ a high degree
of kinetic isotropy can only be obtained by employing models which
locally restore Translational Invariance (TI) in the local direction
of interface motion. The global restoration of TI in fixed directions
provides substantial kinetic anisotropies already at dimensionless
profile resolutions of $\tilde{\width}=1.0$, as shown in Fig.~\ref{fig:Velocity-vs-Orientation}. 
\item The residual anisotropy of the interfacial energy is evaluated by
means of a shape relaxation simulation of one initially cubic particle
in a system under the constraint of a constant particle volume. Fig.~\ref{fig:evaluation-sphericity}
shows the evaluation of the sphericity of the quasi equilibrium particle
shapes as a function of the phase-field profile resolution for different
phase-field models. In any case, the different sharp phase-field models
provide substantially lower energetic anisotropies as compared to
the conventional CF-model. However, for profile resolutions below
$\tilde{\width}<1.3$ grid pinning is observed in unlucky cases using
models with a global restoration of TI in fixed lattice directions.
The TI$_{\mv{n}}$-model reliably provides very small residual interface
energy anisotropies.
\end{itemize}

\authorcontributions{ Conceptualization, M.F. Model \& Code development,
M.F., F.S. and P.Z.; Investigation, M.F., F.S. and P.Z.; Validation,
M.F., F.S. and P.Z.; Writing---original draft preparation, M.F.;
writing-{}-{}-review and editing, M.F., F.S. and P.Z.; Supervision,
M.F.; Funding Acquisition, M.F.}

\funding{The work is funded by the Deutsche Forschungsgemeinschaft
(DFG) -- 431968427.}

\dataavailability{The research data sets generated during the current
study are available from the corresponding author upon reasonable
request.} 

\acknowledgments{We thank A.~Finel from ONERA, Ch\^atillon, France
for fruit-full discussions on this issue.}

\conflictsofinterest{The authors have no conflict of interest to
declare.} 

\abbreviations{Abbreviations}{ The following abbreviations are used
in this manuscript:\\
\begin{tabular}{ll}
SPFM & Sharp Phase-Field Method\tabularnewline
CF & Continuum Field\tabularnewline
TI & Translational Invariance\tabularnewline
FD & Finite Difference\tabularnewline
FFT & Fast Fourier Transformation\tabularnewline
\end{tabular} }

\appendixtitles{no}\appendixstart \appendix\section[\appendixname~\thesection]{} 

Supplementary information is available for this manuscript:
\begin{enumerate}
\item \texttt{Supplementary\_\-material\_\-1\_\-Pinning\_\-animation\-.mpg}:
This movie is an animated version of Figure \ref{fig:pinning}. It
illustrates the influence of spurious grid friction and grid pinning
on the motion of a planar interface in one dimension. We compare the
behavior of the conventional phase-field formulation for different
phase-field widths $\width/\dx$ with the behavior of the sharp phase-field
model (green curves).
\item \texttt{Supplementary\_\-material\_\-2\_\-Steady\_\-state\_\-interface\_\-motion\-.mpg}:
This movie is an animated visualization of the evolution of the phase-field
during the simulation of stationary motion of a planar interface with
propagates under an angle of $\vartheta_{[011]}=30^{\circ}$ with
respect to the computational grid using the TI$_{\mv{n}}$ model with
$\tilde{\width}=0.6$.
\item \texttt{Supplementary\_\-material\_\-3\_\-Contact\-Angle\-BC\-.mpg}:
This movie illustrates the function of the newly proposed boundary
conditions. They enforce a finite contact angle between the interface
normal and the boundary plane. In this movie, we show a simulation
of the shape-evolution of an initially cubic particle toward its spherical
equilibrium shape under conserved phase volume. The particle is in
contact with the bottom boundary, where a contact angle of 80° with
respect to the boundary plane is enforced.
\end{enumerate}
\begin{adjustwidth}{-\extralength}{0cm}

\reftitle{References}

\bibliography{Literature-2Phase}

\end{adjustwidth}
\end{document}